# Quantum Phase Imaging using Spatial Entanglement

**Chien-Hung Lu, Matthew Reichert, Xiaohang Sun, and Jason W. Fleischer**[*]

*Department of Electrical Engineering, Princeton University, Princeton, NJ, 08544, USA*
*Corresponding author: jasonf@princeton.edu

**Abstract**

Entangled photons have the remarkable ability to be more sensitive to signal and less sensitive to noise than classical light. Joint photons can sample an object collectively, resulting in faster phase accumulation and higher spatial resolution[1-3], while common components of noise can be subtracted. Even more, they can accomplish this while physically separate, due to the nonlocal properties of quantum mechanics. Indeed, nearly all quantum optics experiments rely on this separation, using individual point detectors that are scanned to measure coincidence counts and correlations[4]. Scanning, however, is tedious, time consuming, and ill-suited for imaging. Moreover, the separation of beam paths adds complexity to the system while reducing the number of photons available for sampling, and the multiplicity of detectors does not scale well for greater numbers of photons and higher orders of entanglement. We bypass all of these problems here by directly imaging collinear photon pairs with an electron-multiplying CCD camera. We show explicitly the benefits of quantum nonlocality by engineering the spatial entanglement of the illuminating photons and introduce a new method of correlation measurement by converting time-domain coincidence counting into spatial-domain detection of selected pixels. We show that classical transport-of-intensity methods[5] are applicable in the quantum domain and experimentally demonstrate nearly optimal (Heisenberg-limited) phase measurement for the given quantum illumination[6]. The methods show the power of direct imaging and hold much potential for more general types of quantum information processing and control.

While quantum optics in general has gone through several stages of maturation, quantum imaging remains in its infancy. There are two main reasons for this: the relatively recent development of electron-multiplying CCD (EMCCD) cameras suitable for direct imaging, and legacy methods of quantum measurement based on scanning point detectors[7]. In these methods, the emphasis has been on binary polarization orientation for information qubits and action-at-a-distance encryption[8], with little consideration of spatial modes or their propagation. Even with EMCCDs, imaging to date has concentrated on direct intensity measurement in ghost-type geometries, in which one photon in an entangled pair samples the object while the other is detected[9,10]. This separation halves the photon budget for sampling, which is then further reduced by absorption by the object. Moreover, direct intensity measurements limit access to higher orders of quantum coherence[11].

Here, we use a collinear geometry to image a pure phase object with entangled photon pairs. This geometry has been used to directly image spatial Einstein-Podolsky-Rosen (EPR) entanglement[12,13] and allows all the illumination photons to be used for sampling. However, imaging a phase mask cannot be done directly; phase objects give no intensity variation in the focal plane and therefore require either interference or some type of phase retrieval algorithm. Both measurement methods are subtle in the quantum domain, as there is no true phase operator for photons[14], and the fixed number of photons (pairs in this case) guarantees maximal phase uncertainty. On the other hand, phase accumulates during propagation, meaning that phase can be retrieved through a series of intensity measurements[15]. Previous experiments on entangled-pair imaging have used coincidence counting using two separate single-pixel detectors, relying on scanning to determine spatial variation[15-17]. However, scanning is an inefficient method that does not scale well with higher numbers of entangled photons, spatial modes, or measurement planes.

To retrieve the phase, we treat the EMCCD array as a highly parallel multi-pixel detector[18-20] and adapt traditional transport-of-intensity (TIE) methods[5] to the quantum domain. The experimental setup is shown in Fig. 1. As a quantum illumination source, type-I SPDC from a BBO crystal was used to convert a $\lambda_p$= 405 nm laser pump into entangled photon pairs (biphotons) at $2\lambda_p$= 810 nm. The phase object was a letter 'S', placed directly after the output face of the nonlinear crystal. In each exposure, the average photon flux on the camera is about 0.033 photons per pixel, meaning that 1) each pixel is deeply in the quantum regime, 2) the likelihood of a pixel recording a double photon count is extremely low, and 3) images must be built up by summing multiple measurements.

Figures 2a-c show images defocused by 1.6 mm for both classical (coherent-state) and quantum (Fock-state) illumination. To isolate number effects from those of wavelength, we also show results from illumination with an 808nm laser source. This diffracted image is very similar to that of the 405nm pump beam, while the biphoton image is significantly more diffuse (and thus more sensitive to phase). With these images, the phase object can be reconstructed numerically using the TIE equation:

$$-\frac{2\pi}{\lambda I}\frac{\partial I(x,y,z)}{\partial z} = \nabla^2 \varphi(x,y) \qquad (1)$$

where $\varphi(x,y)$ is phase and $I = I(x,y,0)$ is intensity at the focal plane. This equation represents a simple conservation of energy, indicating that variations in intensity along the optical axis can only occur by diffraction in the transverse direction (see Methods). Previously used only in a classical context, its paraxial form is valid here since the scales of the phase object and defocusing distance are both larger (by three orders of magnitude) than the wavelength[4]

Phase reconstruction using Eq. (1) is shown in Figs. 2d-f. Corresponding line profiles, shown in Fig. 2g, compare the statistical results of biphoton imaging with the coherent-state illumination of laser light. In the regions of constant phase, there is significantly less noise in the quantum case; calculating the normalized root-mean-square (RMS) deviation (with respect to wavelength), the two-photon result of 0.14rad gives 150 % and 185 % improvement in the signal-to-noise ratio over classical illumination (0.21 and 0.26 rad for 405nm and 808nm, respectively). This is better than the $\sqrt{2}$ improvement observed in previous ghost-type geometries, where only one of the photons is used for sampling[9,10], and is near the 2× Heisenberg limit predicted for ideal quantum illumination[6]. The result can be improved further, as higher-order TIE[21], higher-order correlations[11,22], and optimal defocusing distance[23] are not accounted for in the measurement and reconstruction algorithm.

To confirm that the quantum nature of the biphoton is responsible for the improved image quality, we perform a second experiment in which the spatial entanglement is engineered to be zero. Just as classical correlations can change upon propagation, e.g. via simple magnification, quantum entanglement can evolve[4,24]. Here, we take advantage of a crossover in spatial entanglement, from full correlation in the near field (at the origin of SPDC) to full anti-correlation in the far field[25], to engineer the degree of coherence. More specifically, by operating at the zero point, we can directly compare the results of quantum illumination with and without spatial entanglement (Fig. 3).

Interestingly, there is no difference in the defocused intensity between the two sources. This is because intensity is a first-order coherence while entanglement is a second-order coherence. Indeed, observations such as this led to the original proposal of higher-order coherence[11] and the need for coincidence counting to measure them[26]. However, the long exposure time necessary to accumulate photons in the EMCCD camera precludes straightforward time gating (as used in intensified cameras[18-20]). Instead, we introduce a purely spatial method based on statistics: we angle the crystal so that the mean spread of the photon pair equals the spacing between nearest-neighbor pixels on the CCD array. If two such pixels are triggered, then it is much more likely that they were illuminated by a biphoton in a single event than by two individual photons in different events. That is, we have converted temporal coincidence into spatial localization (at the expense of significant intensity filtering). Comparing the results with and without entangled illumination (Figs. 3e,f) shows that photon correlations are necessary for improved phase retrieval.

To see more clearly how spatial entanglement manifests itself, we perform direct numerical simulation of the photon propagation (see Methods). In the detection plane, we measure two different quantities with the camera: 1) many short exposures, from which we post-select pixels that correspond to a photon pair, and 2) a long exposure (or, equivalently, summation of the many short exposures without post-selection). These give different measures of the biphoton wavefunction $\psi^{(2)}(x_s, x_i; z)$: 1) a conditional measurement, whose resulting image is proportional to $|\psi^{(2)}(x_s, x_i \approx x_s; z)|^2$, and 2) an unconditional measurement, proportional to $\int |\psi^{(2)}(x_s, x_i; z)|^2 dx_i$.

The results are shown in Fig. 4. The unconditional (integrated) measurement has lower contrast than the conditional measurement (0.73 vs. 0.93), but it also has less fluctuations (RMS of 1.3% vs. 8.7%), as the long exposure time integrates over $x_i$ and averages the effects of interference. In both cases, the variance is considerably less than with classical illumination (21% at 405nm and 15% at 810nm, see Figs. 4a,b), at the price of less-than-perfect contrast. When the quantum illumination is changed, viz. from full to zero spatial entanglement, there is almost no change in the unconditional (integrated) intensity pattern but significant change in the conditional (correlated) measurement (Figs. 4e,f). As in the experiment, removing the correlations at the input erases the information at the output.

The presence of quantum beating leads to two interesting interpretations of the dynamics. First, there can be a superposition between diffraction from the phase object and undisturbed propagation. Second, the components of the biphoton can experience different phase shifts due to sampling different spatial locations on the object. The former is a quantum generalization of self-referenced holography, while the latter is a polarization-free version of differential interference contrast.

Both correlations and which-way ambiguity are necessary to exploit quantum information[14,27]. For the case of two photons, a simple mathematical argument highlights the ambiguity. Labeling the spatial positions of each biphoton component as $x_s$ and $x_i$, consider the mean and difference coordinates $\bar{x} = (x_s + x_i)/2$ and $\Delta x = (x_s - x_i)$. In one interpretation, $\bar{x}$ causes the biphoton to accumulate phase twice as fast as a single photon, leading to enhanced diffraction, while $\Delta x$ subtracts the common noise. In another interpretation, the action is opposite: $\Delta x$ reveals phase differences while $\bar{x}$ averages fluctuations. Either way, the result is enhanced sensitivity with reduced noise.

Finally, we emphasize that the phase estimation here is intimately related to the problem of Fourier transforms, as it is wave interference (vs. simple addition of amplitudes) that gives the final form of the complex field[27]. This is seen most clearly in Figs. 4e and 4f, where quantum beating provides extra oscillations (Fourier modes) that result in more accurate representation of the object (sharper corners and steeper edges). Indeed, the advantages of a quantum algorithm over classical computations results almost entirely from the ability to explore all pathways at once and find the optimal phase for the most likely answer[27-29]. The use of a camera provides a new pathway for these algorithms, as it can process higher-dimensional data at the input and directly image the spatially extended modes at the output. With only two photons in the illumination, only the mean

($\bar{x}$) and difference ($\Delta x$) components of the quantum cosine transform were possible. Nevertheless, the imaging geometries, transport algorithms, and entanglement control demonstrated here are straightforward to generalize to higher numbers of photons and higher degrees of coherence.

**FIGURES**

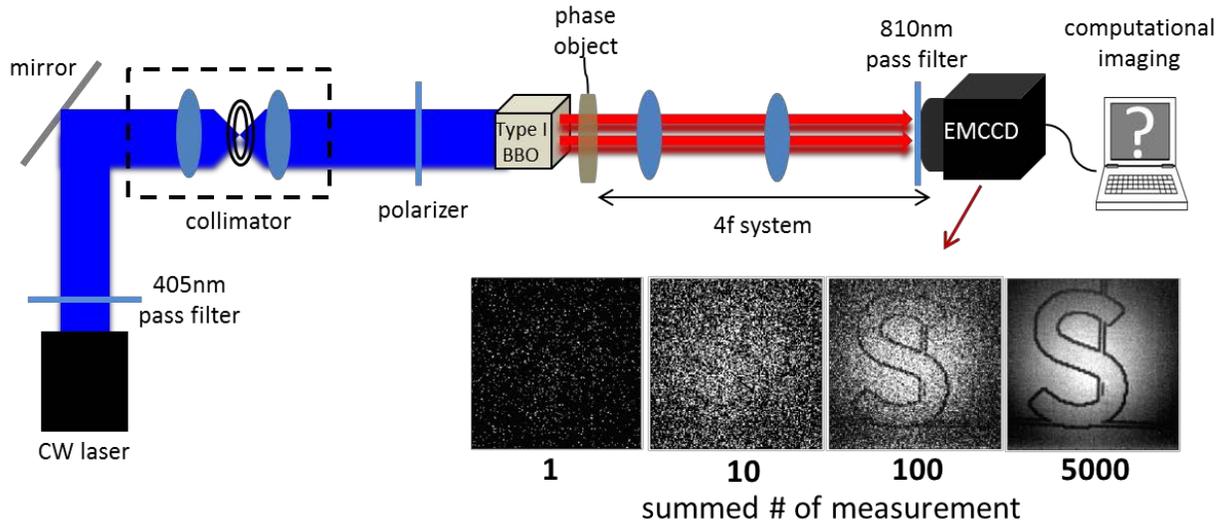

**Figure 1. Schematic of the experiment.** An extraordinarily polarized Gaussian beam (405nm) was sent to a nonlinear crystal BBO. Entangled photon pairs were generated at 810nm by type-I SPDC. A phase object was placed directly after the crystal and photon counting was performed by an electron-multiplying CCD camera. Bottom images show image accumulation after summation of 1, 10, 100 and 5000 frames.

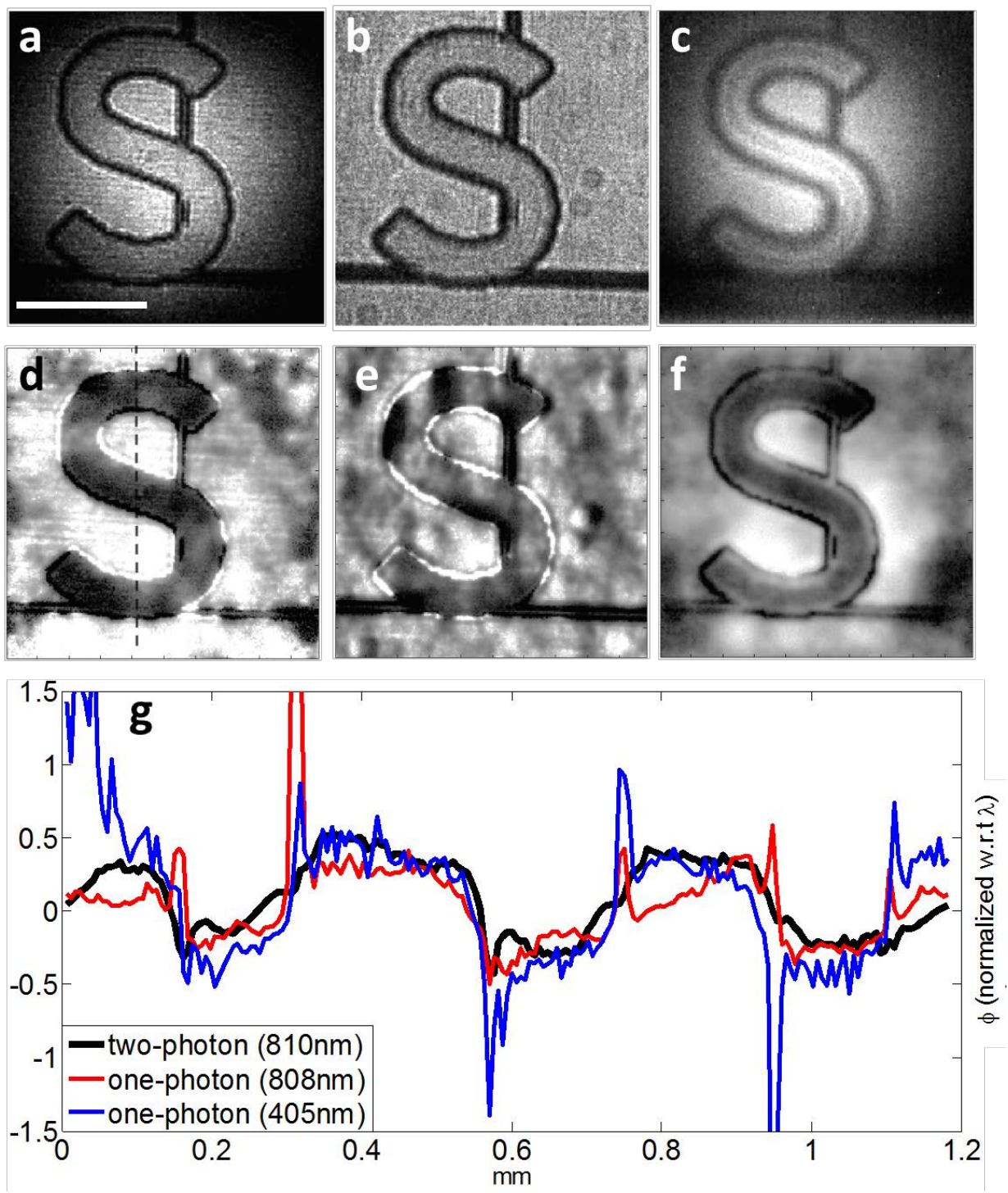

**Figure 2. Experimental results of quantum phase retrieval.** (a-c) Images of the phase object defocused by 1.6mm using (a) 405nm laser light, (b) 808nm laser light, (c) spatially entangled photon pairs at 810nm. (d-f) Numerical reconstructions of phase object using (a-c) in transport-of-intensity equation (1). (g) Line profiles of the dashed cross-section in (d). Scale bar: 0.5 mm

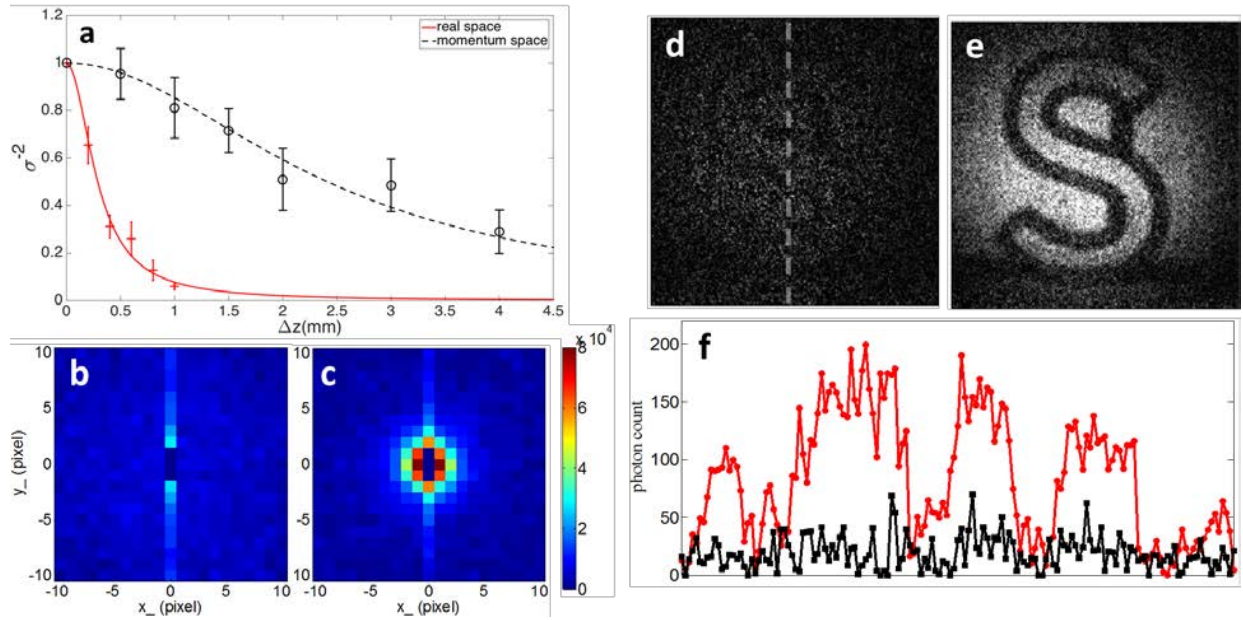

**Figure 3. Spatial localization of coincidence counts.** (a) Fall-off of spatial entanglement with distance from the BBO crystal face. Dots: experiment; lines: fit to Lorentzian decay $\sigma^{-2} = (1 + cz^2)^{-1}$ with distance $z$ from the crystal. (b,c) Two-point correlation measurements without (b) and with (c) spatial entanglement. (d,e) Defocused images of phase object using illumination in (b,c), obtained by summing 100,000 post-selected frames. (f) Average cross-sections of coincidence counts for region outlined in gray dashed line of (d) (black) and (e) (red).

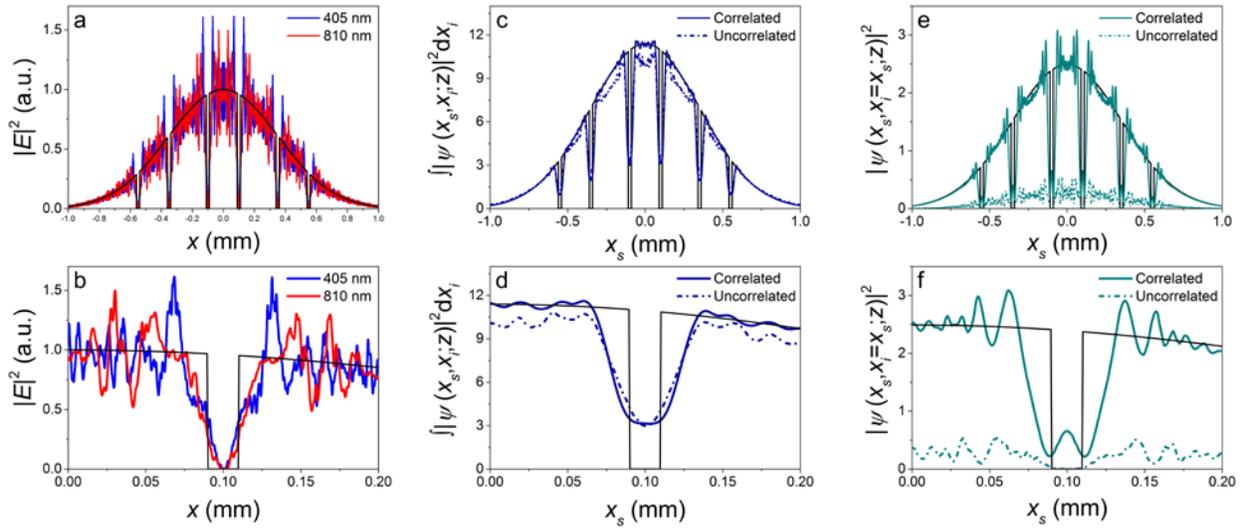

**Figure 4. Numerical simulation of photon propagation.** Top row: defocused images of phase object; bottom row: close-up of central feature. Shown are results for illumination with (a,b) classical (coherent) light and (c-f) photon pairs from SPDC. (c-f) Comparison of quantum results with (solid) and without (dot-dashed) spatially entangled illumination for (c,d) unconditional (integrated) measurement and (e,f) conditional (coincidence) measurement.

## Methods

**Experiment and Verification of Entanglement**

The spatial structure of the two photon field at the crystal can be described by a Gaussian model[13]:

$$\psi(\boldsymbol{\rho}_1, \boldsymbol{\rho}_2) = \frac{1}{\pi\sigma_+\sigma_-} \exp\left[-\frac{|\boldsymbol{\rho}_1+\boldsymbol{\rho}_2|^2}{4\sigma_+^2}\right] \exp\left[-\frac{|\boldsymbol{\rho}_1-\boldsymbol{\rho}_2|^2}{4\sigma_-^2}\right] \quad (2)$$

where $\boldsymbol{\rho}_i = (x_i, y_i)$ is the transverse position of photon $i$ ($i = 1,2$) and $\sigma_+$ and $\sigma_-$ are the strength of the momentum and position correlations, respectively. Experimentally, a 405nm laser pump was collimated into a Gaussian beam with radius $\sigma_p \sim 0.71$mm and sent to a type I BBO crystal. The beam was extraordinarily polarized to maximize the output of spatially entangled photons at 810nm. Position correlation was measured by a 4-f imaging system (lenses with focal lengths of 7.5cm and 20cm, giving a magnification $M = 2.67$). Momentum correlation was measured at the Fourier plane of an optical transform system with an effective focal length $f_e = 12.5$cm (consisting of three lenses with focal lengths 7.5cm, 7.5cm and 12.5cm). For all experiments, the total pump beam power on the front face of the crystal was kept constant at 5mW. Image acquisition was performed by an Andor iXon EMCCD with a pixel size of $16 \times 16$ μm² and quantum efficiency ~70% at 810nm; it was operated at -85 °C, with a region of interest of $201 \times 201$ pixels, a horizontal pixel shift readout rate of 1 MHz, a vertical pixel shift every 0.3μs, a vertical clock amplitude voltage of +4 above the default factory setting, maximal gain, and an exposure time of 0.8ms. The distribution of total noise was characterized in advance, giving a mean of 197 counts and a standard deviation (s.t.d.) of 18 counts. Photon counting was performed by assigning a 1 to pixels with values greater than one s.t.d. above the mean of noise and a 0 otherwise. After acquiring a set of photon counting images, we calculated auto-correlations frame-by-frame and summed all calculations to get the 2D correlation between photon pairs (with contributions from both photon pairs and noise photons). To subtract the contribution from noise photons, a background correlation was obtained by calculating a cross-correlation map using consecutive frames[12,13].

Image appearance after increasing summation of frames is shown in Fig. 1. Position correlation measurements obtained after subtracting the background correlation are shown in Fig. 3c (similar images appear for momentum correlations). Quantitatively, the strength of the momentum and position correlations can be estimated by fitting Gaussians to the joint probabilities, giving $\sigma_+ \approx (\Delta x_+ k\hbar)/f_e = 2.7 \cdot 10^{-3}$ ℏµm$^{-1}$ and $\sigma_- \approx \Delta x_-/M = 9$ µm. Finally, EPR-like correlations can be identified by the inequality:

$$\sigma_+^2 \cdot \sigma_-^2 \cong 5.9 \cdot 10^{-4} \hbar^2 \ll \hbar^2/4 \qquad (3)$$

confirming a significant amount of spatial entanglement.

**Transport-of-intensity Equation**

Equation (1) was implemented by using a finite-difference approximation and a fast Fourier transform (FFT) based solution:

$$\frac{\partial I(x,y,z=0)}{\partial z} \approx \frac{I(x,y,\Delta z) - I(x,y,0)}{\Delta z} \qquad (4)$$

$$\Phi(k_x, k_y) = F(k_x, k_y) \frac{H(k_x,k_y)}{H^2(k_x,k_y)+\gamma_T} \qquad (5)$$

where {$k_x$, $k_y$} are the spatial frequency variables and $\Phi(u,v)$ and $F(u,v)$ are the Fourier transforms of the desired phase $\varphi(x,y)$ and the right hand side of Eq. (4), respectively. In Eq.

(5), $H(k_x, k_y) = -4\pi^2(k_x^2 + k_y^2)$ and $\gamma_T$ is a Tikhonov regularization, used to avoid numerical instability at the origin; its value is taken proportional to the variance of the experimental background noise (5.6 • 10$^{-5}$ for 405nm laser illumination, 1.6 • 10$^{-4}$ for 808nm laser illumination, and 1.6 • 10$^{-5}$ for 810nm biphoton illumination).

**Spatial Localization of Coincidence Counting**

Imaging of two-photon coincidence counts can be obtained by identifying signal photon pairs in each frame. Rotating the BBO crystal allows us to adjust the phase-matching conditions of SPDC and match the mean pair separation distance with the spacing between pixels on the CCD array. At the camera, we examine each frame and keep only nearest-neighbor detections, leading to a first-order removal of isolated (non-coincident) photons. Noise photons are removed further by identifying photons in the central region of the background correlation map, {(x-, y-)| x- = ±1, -1 ≤ y- ≤ +1}, obtained by cross-correlation of consecutive frames. After subtracting noise photons, the signal of photon pairs is maximally optimized, giving an image of coincidence counts by summing over post-selected frames.

**Numerical Simulation of Biphoton Propagation and Imaging**

We are interested in the two-photon state from SPDC, which we assume to be degenerate and collinear ($k_1 = k_2 = k$). In reciprocal space, this state is given by[30]

$$\Phi(q_s, q_i) = \mathcal{N} \operatorname{sinc}\left(\frac{L_z \lambda_p}{8\pi}(q_s - q_i)^2\right) e^{-\sigma_p^2(q_s + q_i)^2}, \qquad (6)$$

where $q_{s(i)}$ is the transverse component of the wavevector for the signal (idler) photon, $\sigma_p$ is the width of the Gaussian pump (at 1/e of irradiance), which has wavelength $\lambda_p$, $L_z$ is the crystal thickness, and $\mathcal{N}$ is a normalization constant

$$\mathcal{N} = \frac{\sqrt{6\sigma_p}}{4}\left(\frac{L\lambda_p}{\pi^3}\right)^{\frac{1}{4}}. \tag{7}$$

The real-space expression, obtained by Fourier transforming Eq. (6), is

$$\psi^{(2)}(x_s, x_i; 0,0) = \frac{\sqrt{2\pi}\mathcal{N}}{\sigma_p L\lambda_p} e^{-\frac{(x_s+x_i)^2}{16\sigma_p^2}} \left[(x_s - x_i)\pi\left(\mathcal{S}\left(\frac{x_s - x_i}{\sqrt{L\lambda_p}}\right) - \mathcal{C}\left(\frac{x_s - x_i}{\sqrt{L\lambda_p}}\right)\right) + \sqrt{L\lambda_p}\left(\cos\left(\frac{\pi}{2L\lambda_p}(x_s - x_i)^2\right) + \sin\left(\frac{\pi}{2L\lambda_p}(x_s - x_i)^2\right)\right)\right], \tag{8}$$

where $x_{s(i)}$ is the transverse signal (idler) coordinate, and $\mathcal{S}$ and $\mathcal{C}$ are Fresnel integrals.

The biphoton wave function evolves upon propagation according to[24]

$$\psi^{(2)}(x'_s, x'_i; z_s, z_i) = \iint \psi^{(2)}(x_s, x_i; 0,0) h_s(x'_s - x_s; z_s) h_i(x'_i - x_i; z_i) \mathrm{d}x_s \mathrm{d}x_i, \tag{9}$$

where $h_{s(i)}$ is the impulse response function of free space propagation given by

$$h(x; z) = \frac{e^{ikz}}{\sqrt{i\lambda z}} e^{i\frac{k}{2z}x^2}. \tag{10}$$

We take $z_s = z_i = z$, such that Eq. (8) represents propagation of the biphoton from an input plane with coordinates $(x_s, x_i)$ to an output plane a distance $z$ away with coordinates $(x'_s, x'_i)$. Equation (9) then becomes

$$\psi^{(2)}(x'_s, x'_i; z) = \frac{e^{i2kz}}{i\lambda z} e^{i\frac{k}{2z}(x'^2_s + x'^2_i)}$$

$$\cdot \iint \psi^{(2)}(x_s, x_i; 0) e^{i\frac{k}{2z}(x^2_s + x^2_i)} e^{i\frac{k}{z}(x_s x'_s + x_i x'_i)} dx_s dx_i,$$

(11)

Equation (11) is the biphoton equivalent of the (1D) classical Fresnel propagation integral (which was used for the classical simulations in Fig. 4.) Finally, the wave function at the detection plane is calculated by propagating Eq. (11) from the crystal to the phase object, multiplying by the phase object's complex transmission function $t(x_s, x_i)$, and propagating again to the detection plane.